%% file: inference_JPCS_final_rev2.tex
\documentclass[a4paper]{jpconf}
\usepackage{graphicx}
\usepackage{amssymb}

\include{commands}
\begin{document}
\title{Inference from correlated patterns: a unified theory for
perceptron learning and linear vector channels}

\author{Yoshiyuki Kabashima}

\address{Department of Computational Intelligence and Systems Science, 
Tokyo Institute of Technology, Yokohama 226-8502, Japan}

\ead{kaba@dis.titech.ac.jp}

\begin{abstract}
A framework to analyze inference performance in 
densely connected single-layer feed-forward networks is 
developed for situations where a given data set 
is composed of correlated patterns. 
The framework is based on the assumption that the left and right 
singular value bases of the given pattern matrix 
are generated independently and uniformly from Haar measures. 
This assumption makes it possible to characterize the objective system by 
a single function of two variables which is determined by the eigenvalue 
spectrum of the cross-correlation matrix of the 
pattern matrix. Links to existing methods for analysis of 
perceptron learning and Gaussian linear vector channels 
and an application to a simple but nontrivial problem are also shown. 
\end{abstract}

\section{Introduction}
Inference from data is one of the most significant problems 
in information science, and perceptrons (or single-layer feed-forward networks)
are often included in widely-used 
devices for solution of this problem. 
In the general scenario, for a given $N$ dimensional input 
pattern $\bx=(x_1,x_2,\ldots,x_N)^{\rm T}$, 
such a network returns an output $y$, which may be a continuous/discrete 
single/multidimensional variable, following 
a conditional probability distribution $P(y|\bx; \bw)=P(y|\Delta)$, 
where ${\rm T}$ denotes the matrix transpose, $\bw=(w_1,w_2,\ldots,w_N)^{\rm T}$ 
denotes the weight parameter of the perceptron and $\Delta=N^{-1/2} \bw \cdot \bx$. 
The scale factor $N^{-1/2}$ is introduced to ensure that 
the components of $\bw$ and $\bx$ are typically of $O(1)$ as 
the limit of $N \to \infty$. 
Given a data set $\xi^p=\{(\bx_1,y_1), (\bx_2,y_2),
\ldots,(\bx_p,y_p)\}$, 
the Bayes formula 
\begin{eqnarray}
P(\bw|\xi^p)=\frac{1}{Z_P(\xi^p)}P(\bw)\prod_{\mu=1}^p P(y_\mu|\Delta_\mu), 
\label{posterior}
\end{eqnarray}
provides us with a useful basis for constructing the optimal inference, 
which may for example involve estimation of the parameter $\bw$, or prediction 
of outputs for novel input patterns. 
Here $P(\bw)$ is a certain prior distribution of $\bw$, 
$\Delta_\mu=N^{-1/2}\bw \cdot \bx_\mu$ $(\mu=1,2,\ldots,p)$ 
and the normalization factor $Z_P(\xi^p)=
\mathop{\rm Tr}_{\bw} P(\bw)\prod_{\mu=1}^p P(y_\mu|\Delta_\mu)$
serves as a partition function, where ${\rm Tr}_{\bw}$ denotes 
summation (or integration) over all possible states of $\bw$. 

In general, equation (\ref{posterior}) can be regarded as
the canonical distribution of a virtual spin system
which is subject to random interactions. 
This similarity has motivated cross-disciplinary 
research across the fields of statistical mechanics and 
neural information processing over the last 
two decades, which has led to the discovery of various 
complex behaviors in the learning processes of neural networks 
\cite{WatkinRauBiehl1992,Engel2001,Nishimori2001}
and to the development of families of advanced mean field 
approximation algorithms
that practically overcome the intrinsic computational difficulties 
underlying inference in large 
networks \cite{AdvancedMeanFieldMethods,MezardZeccina2002}. 

More recently, inference in the style of equation (\ref{posterior}) 
is also being researched actively in another context; namely, in the study of 
linear vector channels for wireless communication. In this 
context, multiple information symbols
denoted by $\bw$ are simultaneously transmitted through a single channel, 
linearly transformed into $\Delta_\mu =N^{-1/2}  \bx_\mu \cdot \bw $ 
$(\mu=1,2,\ldots,p)$. At the receiver's terminal, 
the transmitted symbols $\bw$ have to be estimated from the 
received signals $y_\mu$ $(\mu=1,2,\ldots,p)$. 
Under the assumption that the channel 
and the prior distribution of information symbols 
are modeled as 
$\prod_{\mu=1}^p P(y_\mu|\Delta_\mu)$ and $P(\bw)$, respectively, 
equation (\ref{posterior}) allows  the optimal demodulation scheme. 
The similarity between problems of inference and disordered spin systems
again serves to potentiate nontrivial performance analysis 
\cite{Tanaka2002,GuoVerdu2005,Muller2003,
Wenetal2005,Wen2006,Guo2006,Moustakas2003,TakeuchiTanakaYano2007}
and development of advanced approximate demodulation algorithms 
\cite{Kabashima2003,NeirottiSaad2005,TanakaOkada2005,MontanariPradhakarTse2005,MontanariTse2006} for large systems. 

Although statistical mechanical schemes have been 
applied successfully to various inference problems 
of the form of equation (\ref{posterior}) in such ways, 
there still remain several research directions to explore.
Investigation of inference from {\em correlated patterns}
is a typical example of such a problem. 
For theoretical simplicity, most existing research on perceptron 
learning is based on the assumption that the input vectors are 
independently generated from an isotropic 
distribution \cite{WatkinRauBiehl1992,Engel2001,Nishimori2001}. 
However, it is obvious that real world data is usually 
somewhat biased and correlated across components, which 
makes it difficult to utilize the developed schemes 
directly for data analysis beyond a conceptual level. 
Exploration of correlated patterns is also important 
in the study of linear vector channels because the matrix entries 
of the linear transformation 
are generally correlated with each other due to spatial 
proximity of antennas and for optimizing communication performance
\cite{Verdu1998,TulinoVerdu2004}. Recently, the author and his colleagues 
have developed a framework to handle such situations based on a formula of 
random matrix theory \cite{TakedaUdaKabashima2006,TakedaHatabuKabashima2007}. 
However, the scheme we have developed is still not fully satisfactory because
it is applicable only to Gaussian channels. 
In order to deal with more general situations, further development is required.

The purpose of this article is to provide such a development. 
More precisely, we will develop a framework to analyze 
inference offered by equation (\ref{posterior})
when entries of the pattern matrix 
\begin{eqnarray}
X=N^{-1/2}(\bx_1,\bx_2,\ldots,\bx_p)^{\rm T}, 
\label{patternmatrix}
\end{eqnarray}
are correlated.
A similar direction has already been followed by Opper and Winther 
\cite{OpperWinther2001L,OpperWinther2001,OpperWinther2005}. 
However, their formalism, developed for densely connected networks
of two-body interactions, is highly general, and therefore 
properties that hold specifically for models satisfying equation 
(\ref{posterior}) are not fully utilized. 
Hence we develop here a specific formalism for analyzing 
inference problems expressed by means of equation (\ref{posterior}). 

This article is organized as follows. 
In section 2, models that we will investigate are introduced. 
For characterizing correlated patterns, 
we assume that the pattern matrix (\ref{patternmatrix}) is 
randomly generated under the constraint that singular values of 
the matrix obey a given distribution. 
Section 3 is the main part of this article, 
in which two analytical schemes are developed. 
One takes as its basis the replica method \cite{Dotzenko2001}, 
which makes it possible to assess the typical inference 
performance of the objective system by averaging the pattern matrix $X$ 
with respect to an assumed distribution. 
The other is developed for approximately evaluating 
the average of $\bw$ with respect to equation (\ref{posterior})
for a given specific $X$ (or $\xi^p$), which corresponds
to the Thouless-Anderson-Palmer approach \cite{TAP1978} 
in spin glass research. It is shown that 
a two-variable function, which we denote by $F(x,y)$ and which
is determined by the eigenvalue spectrum of the cross-correlation matrix 
$X^{\rm T}X$ and the pattern ratio $\alpha=p/N$, 
plays an important role in both schemes. 
Links to existing methods of analysis of the schemes 
that we develop are indicated 
in section 5 in conjunction with an application to a 
simple example problem. The final section contains a summary. 

\section{Model definition}
An expression of the singular value decomposition 
\begin{eqnarray}
X=U^{\rm T}DV, 
\label{singularvalue}
\end{eqnarray}
of the pattern matrix $X$ is the basis of our framework, where 
$D={\rm diag}(d_k)$ is a $p \times N$ diagonal matrix, and 
$U$ and $V$ are $p \times p$ and $N \times N$ orthogonal 
matrices, respectively. 
Linear algebra guarantees that an 
arbitrary $p \times N$ matrix $X$ can be decomposed 
according to equation (\ref{singularvalue}). 
The singular values of $X$, $d_k$ $(k=1,2,\ldots, {\rm min}(p,N))$, 
are linked to eigenvalues of the cross correlation $X^{\rm T} X$, 
$\lambda_k$ $(k=1,2,\ldots,N)$, as
$\lambda_k=d_k^2$ $(k=1,2,\ldots, {\rm min}(p,N))$ and 
$0$ otherwise, where ${\rm min}(p,N)$ denotes the 
lesser value of $p$ and $N$. 
In order to handle correlations in $X$ analytically, we 
assume that the orthogonal matrices $U$ and $V$ are uniformly 
and independently generated from the Haar measures of 
$p \times p$ and $N \times N$ orthogonal matrices, respectively, 
and that the empirical eigenvalue spectrum  of $X^{\rm T}X$,
$N^{-1}\sum_{k=1}^N \delta(\lambda-\lambda_k)=
(1-{\rm min}(p,N)/N) \delta(\lambda)+ 
N^{-1} \sum_{k=1}^{{\rm min}(p,N)} \delta(\lambda-d_k^2)$, 
converges to a certain distribution $\rho(\lambda)$
as $N$ and $p$ tend to infinity with keeping $\alpha =p/N$ 
of the order of unity. 

For generality, we assume that the outputs $\by=(y_1,y_2,\ldots,y_p)^{\rm T}$
for $X$ are generated from a {\em generative model }
\begin{eqnarray}
Q(\by|X)=\mathop{\rm Tr}_{\bw} Q(\bw) \prod_{\mu=1}^p Q(y_\mu|\Delta_\mu)
=Z_Q(\xi^p), 
\label{generative}
\end{eqnarray}
where the prior and conditional probabilities of this model, 
$Q(\bw)$ and $Q(y|\Delta)$, may differ from those of the {\em recognition model}, 
$P(\bw)$ and $P(y|\Delta)$, which is used in equation (\ref{posterior}). 
For a fixed data set $\xi^p=\{(\bx_1,y_1),
(\bx_2,y_2),\ldots,(\bx_p,y_p)\}=(X,\by)$, $Q(\by|X)=Z_Q(\xi^p)$ serves as 
the partition function of the correct posterior distribution of 
$\bw$, $Q(\bw|\xi^p)=Q(\bw) \prod_{\mu=1}^p Q(y_\mu|\Delta_\mu)/Z_Q(\xi^p)$. 
For analytical tractability, we also assume that both the 
prior distributions of the generative and recognition models 
can be factorized as
$Q(\bw)=\prod_{i=1}^N Q(w_i)$ and $P(\bw)=\prod_{i=1}^N P(w_i)$, respectively.

\section{Analysis}
\subsection{Analysis of the generative model and the $F$-function}
We first analyze properties of the generative model
since outputs $\by$ of the data set $\xi^p$ are generated by this model 
following equation (\ref{generative}). 
For this purpose, we introduce an expression 
\begin{eqnarray}
Z_Q(\xi^p)&=&
\mathop{\rm Tr}_{\bw}\prod_{i=1}^N Q(w_i) 
\prod_{\mu=1}^p 
\left (
\int d  \Delta_\mu 
Q(y_\mu|\Delta_\mu)\delta(\Delta_\mu-N^{-1/2}\bw \cdot \bx_\mu)\right ) \cr
&=& \int \prod_{\mu=1}^p 
\left (\frac{d u_\mu d \Delta_\mu }{2 \pi} 
\exp \left [-{\rm i} u_\mu \Delta_\mu \right ] Q(y_\mu|\Delta_\mu) \right )
\mathop{\rm Tr}_{\bw}\prod_{i=1}^N Q(w_i) \exp \left [{\rm i} \bu^{\rm T}X \bw 
\right ] \cr
&=& \mathop{\rm Tr}_{\bu,\bw} \prod_{\mu=1}^p \widehat{Q}_{y_\mu}(u_\mu) 
\prod_{i=1}^N Q(w_i) \exp \left [{\rm i} \bu^{\rm T}X \bw 
\right ],
\label{generative_partition} 
\end{eqnarray}
where ${\rm i}=\sqrt{-1}$, $\bu=(u_1,u_2,\ldots,u_p)^{\rm T}$ and 
$\widehat{Q}_{y_\mu}(u_\mu)=\int d\Delta_\mu
\exp \left [-{\rm i} u_\mu \Delta_\mu \right ] Q(y_\mu|\Delta_\mu) /(2 \pi)$. 
Next, we substitute equation (\ref{singularvalue}) into 
equation (\ref{generative_partition}) and take an average with respect to the
orthogonal matrices $U$ and $V$. 
For this evaluation, it is noteworthy that for {\em fixed} sets 
of dynamical variables
$\bw$ and $\bu$, $\widetilde{\bw}=V\bw$ and $\widetilde{\bu}=U\bu$ behave as 
continuous random variables which are uniformly generated under the
strict constraints
\begin{eqnarray}
&& \frac{1}{N}|\widetilde{\bw}|^2 =\frac{1}{N} |\bw|^2 = T_w, \label{w_norm} \\
&& \frac{1}{p}|\widetilde{\bu}|^2 =\frac{1}{p} |\bu|^2 = T_u, \label{u_norm} 
\end{eqnarray}
when $U$ and $V$ are independently and 
uniformly generated from the Haar measures. 
In the limit as $N,p \to \infty$ with keeping $\alpha=p/N$ $O(1)$, 
this yields an expression 
\begin{eqnarray}
\frac{1}{N} \ln \left [\overline{\exp \left [{\rm i} \bu^{\rm T}X \bw  
\right ]} \right ]=F(T_w,T_u), 
\label{generative_F}
\end{eqnarray}
where $\overline{\cdots}$ denotes 
averaging with respect to the Haar measures, 
the function $F(x,y)$ is defined as
\begin{eqnarray}
&&F(x,y)=\mathop{\rm Extr}_{\Lambda_x,\Lambda_y}
\left \{-
\frac{1}{2}\left \langle \ln (\Lambda_x \Lambda_y+\lambda)
\right \rangle_\rho 
-\frac{\alpha-1}{2}\ln \Lambda_y+\frac{\Lambda_x x}{2}
+\frac{\alpha \Lambda_y y}{2}\right \} \cr
&&\phantom{F(x,y)=}
-\frac{1}{2}\ln x -\frac{\alpha}{2}\ln y-\frac{1+\alpha}{2}, 
\label{F_func}
\end{eqnarray}
and $\left \langle \cdots \right \rangle_\rho$ indicates 
averaging with respect to the asymptotic eigenvalue spectrum of 
$X^{\rm T}X$, $\rho(\lambda)$. 
$\mathop{\rm Extr}_{\theta} \left \{ \cdots \right \}$ 
represents 
extremization with respect to $\theta$, 
which corresponds to the saddle point assessment of a
complex integral and therefore does not necessarily 
mean operation of minimum or maximum. Expressions analogous to 
equations (\ref{generative_F}) and (\ref{F_func}) are
known as the Itzykson-Zuber integral or $G$-function 
for ensembles of square (symmetric) matrices 
\cite{ItzyksonZuber1980,MarinariParisiRitort1994,ParisiPotters1995,
CherrierDeanLefevre2003}. 
These equations imply that the annealed average of 
equation (\ref{generative_partition}) is evaluated as 
\begin{eqnarray}
\frac{1}{N}\ln \left [
\mathop{\rm Tr}_{\by} 
\overline{Z_Q(\xi^p) }
\right ]=\mathop{\rm Extr}_{T_w,T_u}
\left \{F(T_w,T_u)+A_w(T_w)+\alpha A_u(T_u) \right \}, 
\label{generative_action}
\end{eqnarray}
where 
\begin{eqnarray}
A_w(T_w)=\mathop{\rm Extr}_{\widehat{T}_w}
\left
\{ \frac{\widehat{T}_w T_w}{2} + 
\ln \left [\mathop{\rm Tr}_{w} Q(w)\exp \left [-\frac{\widehat{T}_w}{2}w^2 \right ]
\right ]
\right \}, 
\label{AS}
\end{eqnarray}
\begin{eqnarray}
A_u(T_u)=\mathop{\rm Extr}_{\widehat{T}_u}
\left
\{ \frac{\widehat{T}_u T_u}{2} + 
\ln \left [\mathop{\rm Tr}_{u,y} 
\widehat{Q}_y(u)\exp \left [-\frac{\widehat{T}_u}{2}u^2 \right ]
\right ]
\right \}. 
\label{Au}
\end{eqnarray}
Normalization constraints $\mathop{\rm Tr}_y Q(y|\Delta)=1$ 
guarantee that ${\rm Tr}_{\by} \overline{Z_Q(\xi^p)}=1$, 
which, in conjunction with equations (\ref{generative_action}), 
(\ref{AS}) and (\ref{Au}), 
implies that $T_w={\rm Tr}_w w^2 Q(w)$, $T_u=0$, 
$\widehat{T}_w=0$ and $\widehat{T}_u= 
\alpha^{-1} T_w \left \langle \lambda \right \rangle_\rho$.  
The physical implication is that, due to the central limit theorem, 
$\bDelta=(\Delta_1,\Delta_2,\ldots,\Delta_p)^{\rm T}$
follows an isotropic Gaussian distribution
\begin{eqnarray}
Q(\bDelta)=\frac{1}{(2 \pi \widehat{T}_u)^{p/2}} 
\exp \left [-\frac{|\bDelta|^2}{2\widehat{T}_u  } \right ]
= \frac{\alpha^{p/2}}{(2 \pi T_w \left \langle \lambda \right \rangle_\rho
)^{p/2}}
\exp \left [
-\frac{\alpha |\bDelta|^2}{2 T_w \left \langle \lambda \right \rangle_\rho}
\right ], 
\label{delta_dist} 
\end{eqnarray}
in the limit as $N,p \to \infty$, $\alpha=p/N \sim O(1)$ 
when $\bw$ is generated from $Q(\bw)=\prod_{i=1}^N Q(w_i)$, and 
$U$ and $V$ are independently and uniformly generated from the Haar measures. 

\subsection{Replica analysis}
Now, we are ready to analyze equation (\ref{posterior}). 
As $\xi^p$ is a set of predetermined random variables 
depending on $X$ and the generative model (\ref{generative_partition}), 
we utilize the replica method. 
This means that we evaluate the $n$-th moments of the partition 
function $Z_P(\xi^p)$ for natural numbers 
$n\in \mN$ as
\begin{eqnarray}
\left [ Z_P^n (\xi^p) 
\right ]_{\xi^p}
&=&\mathop{\rm Tr}_{\by}
\overline{
Q(\by|X) Z_P^n (\xi^p) }=\mathop{\rm Tr}_{\by}
\overline{
Z_Q(\xi^p) Z_P^n (\xi^p) } \cr
&=&\mathop{\rm Tr}_{
\{\bu^a\},\{\bw^a\}}
\prod_{\mu=1}^p
\left ({\rm Tr}_{y_\mu}\widehat{Q}_{y_\mu}(u^{0}_\mu)
\prod_{a=1}^n \widehat{P}_{y_\mu}(u^{a}_\mu ) \right )
\times \prod_{i=1}^N 
\left (Q(w_i^0)
\prod_{a=1}^n P(w_i^a) \right ) \cr
&& \phantom{\left [ Z_P^n (\xi^p) 
\right ]_{\xi^p}=}
\times \overline{\exp \left [{\rm i}\sum_{a=0}^n 
(\bu^a)^{\rm T}X \bw^a \right ]}, 
\label{moments_natural}
\end{eqnarray}
and assess the quenched average of free energy with respect to the 
data set $\xi^p$ as 
$ N^{-1}\left [ \ln  Z_P (\xi^p) \right ]_{\xi^p} =\lim_{n \to 0}
\frac{\partial}{\partial n}
N^{-1}
\ln \left [Z_P^n (\xi^p) \right ]_{\xi^p}$,
analytically continuing expressions obtained for
equation (\ref{moments_natural}) from 
$n \in \mN$ to real numbers $n \in \mR$. 
Here, $\left [ \cdots \right ]_{\xi^p}=
\mathop{\rm Tr}_{\by} \overline{Q(\by|X)( \cdots )}
=\mathop{\rm Tr}_{\by} \overline{Z_Q(\xi^p)( \cdots )}$
represents the average with respect to the data set $\xi^p$, 
$\widehat{P}_{y_\mu}(u_\mu)=\int d\Delta_\mu
\exp \left [-{\rm i} u_\mu \Delta_\mu \right ] P(y_\mu|\Delta_\mu) /(2 \pi)$. 
$\{\bw^a\}$ and $\{\bu^a\}$ represent sets of dynamical 
variables $\bw^0,\bw^1,\ldots,\bw^n$ and
$\bu^0,\bu^1,\ldots,\bu^n$, respectively, 
where the replica indices $0$ and $1,2,\ldots,n$ denote
the generative and $n$ replicas of recognition models, respectively. 

For this procedure, a note similar to that for the evaluation of 
equation (\ref{generative_F}) is useful. Namely, for fixed sets of 
dynamical variables $\{\bu^a\}$ and $\{\bw^a\}$, 
$\widetilde{\bu}^a=U \bu^a$ and $\widetilde{\bw}^a=V \bw^a$ 
behave as continuous random variables which satisfy strict constraints
\begin{eqnarray}
&&\frac{1}{N} \widetilde{\bw}^a \cdot \widetilde{\bw}^b
=\frac{1}{N} {\bw}^a \cdot {\bw}^b =\qw^{ab},  \label{qsab}\\
&&\frac{1}{p} \widetilde{\bu}^a \cdot \widetilde{\bu}^b
=\frac{1}{p} {\bu}^a \cdot {\bu}^b =\qu^{ab}, \label{quab}
\end{eqnarray}
$(a,b=0,1,\ldots,n)$
when $U$ and $V$ are 
independently and uniformly generated from the Haar measures. 
This indicates that equation (\ref{moments_natural}) can be 
evaluated by the saddle point method with respect to 
sets of macroscopic parameters $\cqw=(\qw^{ab})$ and 
$\cqu=(\qu^{ab})$ 
in the limit as $N,p \to \infty$, $\alpha=p/N \sim O(1)$. 
In addition, intrinsic permutation symmetry among replicas indicates
that it is natural to assume that 
$(n+1) \times (n+1)$ matrices $\cqw$ and $\cqu$ are of the form 
\begin{eqnarray}
\cqw&=&\left (
\begin{array}{cccccc}
T_w   & \vline& m_w    &m_w    & \ldots & m_w \cr
\hline 
m_w  & \vline & \cw+q_w &q_w    & \ldots & q_w \cr
m_w  & \vline& q_w    &\cw+q_w & \ldots & q_w \cr
\vdots& \vline & \vdots  &\vdots  & \ddots & \vdots \cr
m_w  & \vline & q_w    &q_w    & \ldots & \cw +q_w 
\end{array}
\right )\cr
&=& E \times \left (
\begin{array}{ccccccc}
T_w            & \sqrt{n} m_w &   \vline  & 0 & 0 & \ldots & 0 \cr
\sqrt{n} m_w  & \cw+nq_w      &   \vline  & 0 & 0 & \ldots & 0 \cr
\hline 
0              & 0             &   \vline  & \cw & 0& \ldots & 0 \cr
0              & 0             &   \vline  &  0  & \cw & \ldots & 0 \cr
\vdots    & \vdots        &   \vline  & \vdots & \vdots & \ddots &\vdots\cr
0         & 0        &   \vline  & 0 & 0 & \ldots & \cw
\end{array}
\right ) \times E^{\rm T}, 
\label{Qs}
\end{eqnarray}
and 
\begin{eqnarray}
\cqu&=&\left (
\begin{array}{cccccc}
T_u   & \vline& -m_u    &-m_u    & \ldots & -m_u \cr
\hline 
-m_u  & \vline & \cu-q_u &-q_u    & \ldots & -q_u \cr
-m_u  & \vline& -q_u    &\cu-q_u & \ldots & -q_u \cr
\vdots& \vline & \vdots  &\vdots  & \ddots & \vdots \cr
-m_u  & \vline & -q_u    &-q_u    & \ldots & \cu -q_u 
\end{array}
\right ) \cr
&=&E \times \left (
\begin{array}{ccccccc}
T_u            & -\sqrt{n} m_u &   \vline  & 0 & 0 & \ldots & 0 \cr
-\sqrt{n} m_u  & \cu-nq_u      &   \vline  & 0 & 0 & \ldots & 0 \cr
\hline 
0              & 0             &   \vline  & \cu & 0& \ldots & 0 \cr
0              & 0             &   \vline  &  0  & \cu & \ldots & 0 \cr
\vdots    & \vdots        &   \vline  & \vdots & \vdots & \ddots &\vdots\cr
0         & 0        &   \vline  & 0 & 0 & \ldots & \cu
\end{array}
\right ) \times E^{\rm T}, 
\label{Qu}
\end{eqnarray}
at the saddle point. Here, 
$E=(\be_0,\be_1,\ldots,\be_n)$ denotes an $n+1$-dimensional
orthonormal basis composed of 
$\be_0=(1,0,0,\ldots,0)^{\rm T}$, $\be_1=(0,n^{-1/2},n^{-1/2},\ldots,
n^{-1/2})^{\rm T}$ and $n-1$ orthonormal vectors $\be_2, \be_3,\ldots, \be_n$, 
which are orthogonal to both $\be_0$ and $\be_1$. 
Rather laborious but straightforward calculation 
on the basis of expressions (\ref{Qs}) and (\ref{Qu}) yields
\begin{eqnarray}
&&\lim_{n \to 0}\frac{\partial}{\partial n} \frac{1}{N}\ln 
\left [\overline{\exp \left [{\rm i}\sum_{a=0}^n 
(\bu^a)^{\rm T}X \bw^a \right ]} \right ]
=\cA_0(\cw,\cu,\qw,\qu,m_w,m_u)\cr
&& =F(\cw,\cu)\!+\!\qw \!\frac{\partial F}{\partial \cw}\!-\!
\qu \!\frac{\partial F}{\partial \cu} 
\!+\! T_w \!\left (\frac{m_u}{\cu} \right )^2\!
\left ( \! \frac{\left \langle \lambda \right \rangle_\rho \cu}{2}
\!+\! \frac{\partial F}{\partial \cw} \right )
\!-\! 2 m_w \left (\frac{m_u}{\cu} \right ) 
\! \frac{\partial F}{\partial \cw}, 
\label{A0}
\end{eqnarray}
where $T_w=\mathop{\rm Tr}_{w} w^2 Q(w)$. 
This equation and evaluation of the volumes of dynamical variables 
$\{\bw^a\}$ and $\{\bu^a\}$ under constraints (\ref{qsab}) and (\ref{quab}) 
of the replica symmetric (RS) ansatz (\ref{Qs}) and (\ref{Qu}) 
provide an expression for the average free energy 
\begin{eqnarray}
&& \frac{1}{N}\left [ \ln  Z_P (\xi^p) \right ]_{\xi^p} =
\lim_{n \to 0} \frac{\partial}{\partial n} \frac{1}{N}
\ln \left [  Z_P^n (\xi^p) \right ]_{\xi^p} \cr
&& =\mathop{\rm Extr}_{\bTheta}
\left \{
\cA_0(\cw,\cu,\qw,\qu,m_w,m_u)+
\cA_w(\cw,\qw,m_w)+
\alpha \cA_u(\cu,\qu,m_u) 
\right \}, 
\label{replica_free_energy}
\end{eqnarray}
where $\bTheta=(\cw,\cu,\qw,\qu,m_w,m_u)$, 
\begin{eqnarray}
&&\cA_w(\cw,\qw,m_w)
=\mathop{\rm Extr}_{\hcw,\hqw,\widehat{m}_w}
\left \{
\frac{\hcw}{2}(\cw+\qw)-\frac{\hqw}{2}\cw-\widehat{m}_w m_w \right .\cr
&& 
\left . \phantom{\cA_w}
+\mathop{\rm Tr}_{w^0} Q(w^0) \int Dz \ln \left [
\mathop{\rm Tr}_w P(w) 
\exp \left [
-\frac{\hcw}{2}w^2+(\sqrt{\hqw}z+\widehat{m}_w w^0)w \right ]
\right ]
\right \}, 
\label{cAs}
\end{eqnarray}
and 
\begin{eqnarray}
&&\cA_u(\cu,\qu,m_u)
=\mathop{\rm Extr}_{\hcu,\hqu,\widehat{m}_u}
\left \{
\frac{\hcu}{2}(\cu-\qu)+\frac{\hqu}{2}\cu-\widehat{m}_u m_u \right .\cr
&& 
\left . \phantom{\cA_w}
+\mathop{\rm Tr}_y \int Dz Dx Q \left (y|\sqrt{\widehat{T}_u-
\frac{\widehat{m}_u^2}{\hqu}}x+\frac{\widehat{m}_u}{\sqrt{\hqu}} z  \right ) 
\ln \left [
\int Dx P(y|\sqrt{\hcu} x+\sqrt{\hqu} z)
\right ]
\right \}. 
\label{cAu}
\end{eqnarray}
Here, $\widehat{T}_u=\alpha^{-1} T_w \left \langle \lambda \right \rangle_\rho$
and $Ds=ds \exp \left [-s^2/2 \right ]/\sqrt{2 \pi}$ represents the Gaussian 
measure. 
Expressions (\ref{A0})--(\ref{cAu}) are the main results of this 
article. 

Two points are noteworthy here. The first is that 
a set of parameters $\bTheta$ determined by the extremizing 
equation (\ref{replica_free_energy}) represents 
typical macroscopic averages of the posterior distribution
(\ref{posterior}), by which various performance measures 
can be evaluated \cite{Engel2001}. 
Moreover, equation (\ref{replica_free_energy})
itself is linked to information theoretic measures
for assessing inference performance. 
For example, the Kullback-Leibler divergence (per output) between 
the generative and recognition models, which represents 
a certain distance from the generative model and is related to 
the prediction ability of the recognition model for novel data, 
is evaluated as
\begin{eqnarray}
{KL}(Q|P)=\frac{1}{p} \mathop{\rm Tr}_{\by}
\overline{Q(\by|X)\ln \frac{Q(\by|X)}{P(\by|X)}} 
=\frac{1}{\alpha N}\left [\ln Z_Q(\xi^p) \right ]_{\xi^p}-
\frac{1}{\alpha N}\left [\ln Z_P(\xi^p) \right ]_{\xi^p}, 
\label{KLD}
\end{eqnarray}
utilizing equation (\ref{replica_free_energy}) \cite{GyorgyiTishby1990}. 
Equation (\ref{replica_free_energy}), in conjunction with 
equation (\ref{delta_dist}), can also be used for 
calculating the typical mutual information (per output) between the parameter $\bw$ and 
the output $\by$, which represents the information content of 
$\bw$ that can be gained by observing the output $\by$
for typical pattern matrices $X$, as
\begin{eqnarray}
I(W;Y)&=&\frac{1}{p}
\mathop{\rm Tr}_{\bw,\by}
Q(\bw) 
\overline{\left [\prod_{\mu=1}^p Q(y_\mu|\Delta_\mu)\right ]
\ln \left [\prod_{\mu=1}^p Q(y_\mu|\Delta_\mu)\right ]
}
-\frac{1}{p}\mathop{\rm Tr}_{\by} \overline{Q(\by|X) \ln Q(\by|X)} \cr
&=& 
\mathop{\rm Tr}_{y}
\int Dz Q\left (y|\sqrt{\widehat{T}_u} z \right ) \ln 
Q\left (y|\sqrt{\widehat{T}_u} z \right ) 
- \frac{1}{\alpha N} \left [ \ln Z_Q(\xi^p) \right ]_{\xi^p}, 
\label{MutualInformation}
\end{eqnarray}
specific expressions of which, for problems of communication through 
additive channels, have been derived in earlier 
studies \cite{Tanaka2002,GuoVerdu2005,Muller2003,Tanaka2004}. 
The other issue is that the current formalism can be applied not
only to the RS analysis presented above but also to that 
of replica symmetry breaking (RSB) \cite{MPV1987}. 
Analysis of the local instability condition of 
the RS solution (\ref{Qs}) and (\ref{Qu})
subject to infinitesimal perturbation of the 
form of the one step RSB yields
\begin{eqnarray}
\left (1-2 \frac{\partial^2 F}{\partial \cw^2}\cw^{(2)} \right )
\left (1-\frac{2}{\alpha} 
\frac{\partial^2 F}{\partial \cu^2}\cu^{(2)} \right )
-\frac{4}{\alpha}\left (\frac{\partial^2 F}{\partial \cw \partial \cu}
\right )^2 \cw^{(2)}\cu^{(2)} < 0, 
\label{AT}
\end{eqnarray}
where 
\begin{eqnarray}
\cw^{(2)}=\mathop{\rm Tr}_{w^0}Q(w^0) \int Dz 
\left ( 
\frac{\partial^2}{\partial \left (\sqrt{\hqw}z\right )^2}
\ln \left [
\mathop{\rm Tr}_{w}
P(w) \exp \left [-\frac{\hcw}{2}w^2+(\sqrt{\hqw}z+\widehat{m}_w w^0)w \right ]
\right ]
\right )^2, 
\label{cs2}
\end{eqnarray}
and 
\begin{eqnarray}
&&\cu^{(2)}=\mathop{\rm Tr}_{y} \int Dz Dx Q\left (y| 
\sqrt{\widehat{T}_u-\frac{\widehat{m}_u^2}{\hqu}} x+
\frac{\widehat{m}_u}{\sqrt{\hqu}} z \right ) 
\cr
&&
\phantom{\cu^{(2)}=aaaaaa}
\times 
\left ( \frac{\partial^2}{\partial 
\left (\sqrt{\hqu}z\right )^2}
\ln 
\left [\int Dx P\left (y|\sqrt{\hcu}x+\sqrt{\hqu}z \right ) 
\right ]
\right )^2. 
\label{cu2}
\end{eqnarray}
Equation (\ref{AT}) corresponds to the de Almeida-Thouless (AT) condition 
for the current system \cite{AT1978}. 

\subsection{The Thouless-Anderson-Palmer approach}
The scheme developed so far can be used for 
macroscopically characterizing the inference performance 
of equation (\ref{posterior}) 
for typical samples of $\xi^p$. However, another method
is necessary to  evaluate microscopic averages 
for an individual sample of $\xi^p$. 
The Thouless-Anderson-Palmer (TAP) approach \cite{TAP1978} known 
in spin glass research offers a useful guideline for this purpose. 
Although several formalisms are known for this approximation scheme 
\cite{AdvancedMeanFieldMethods}, we here follow the one based on the Gibbs
free energy because of its generality and wide 
applicability \cite{OpperWinther2005,ParisiPotters1995}. 

Let us suppose a situation for which the microscopic averages 
of the dynamical variables
\begin{eqnarray}
\bm_{w}=\mathop{\rm Tr}_{\bw} \bw P(\bw|\xi^p)  
=\frac{1}{Z_P(\xi^p)}
\mathop{\rm Tr}_{\bu,\bw} \bw \prod_{\mu=1}^p \widehat{P}_{y_\mu}(u_\mu)
\prod_{i=1}^N P(w_i) \exp \left [{\rm i} \bu^{\rm T}X \bw \right ], 
\label{micro_w_av}
\end{eqnarray}
and 
\begin{eqnarray}
\bm_{u}=\frac{1}{Z_P(\xi^p)}
\mathop{\rm Tr}_{\bu,\bw} \left ({\rm i}\bu \right )
\prod_{\mu=1}^p \widehat{P}_{y_\mu}(u_\mu)
\prod_{i=1}^N P(w_i) \exp \left [{\rm i} \bu^{\rm T}X \bw \right ], 
\label{micro_u_av}
\end{eqnarray}
are required. The Gibbs free energy 
\begin{eqnarray}
\Phi(\bm_w,\bm_u)=
\mathop{\rm Extr}_{\bh_w,\bh_u}
\left \{ \bh_w \cdot \bm_w +\bh_u \cdot \bm_u -
\ln \left [
Z_P(\bh_w,\bh_u)
\right]
\right \}, 
\label{gibbs_FE}
\end{eqnarray}
where 
\begin{eqnarray}
Z_P(\bh_w,\bh_u)=
\mathop{\rm Tr}_{\bu,\bw} 
\prod_{\mu=1}^p \widehat{P}_{y_\mu}(u_\mu)
\prod_{i=1}^N P(w_i) \exp \left [
\bh_w \cdot \bw+\bh_u \cdot ({\rm i}\bu)+
({\rm i} \bu)^{\rm T}X \bw \right ],
\label{h_partition}
\end{eqnarray}
offers a useful basis for this objective as the extremization 
conditions of equation (\ref{gibbs_FE}) generally 
agree with equations (\ref{micro_w_av}) and (\ref{micro_u_av}). 
This indicates that one can evaluate the microscopic averages
(\ref{micro_w_av}) and (\ref{micro_u_av}) by extremization 
once the function of Gibbs free energy (\ref{gibbs_FE}) is provided. 

Unfortunately, exact evaluation of equation (\ref{gibbs_FE}) is 
computationally difficult and therefore we resort to 
approximation. For this purpose, we put a parameter $l$ 
in front of $X$ in equation 
(\ref{h_partition}), which yields the generalized Gibbs free energy as
\begin{eqnarray}
\widetilde{\Phi}(\bm_w,\bm_u;l)=\mathop{\rm Extr}_{\bh_w,\bh_u}
\left \{ \bh_w \cdot \bm_w +\bh_u \cdot \bm_u -
\ln \left [Z_P(\bh_w,\bh_u;l)\right]
\right \}, 
\label{gen_gibbs_FE}
\end{eqnarray}
where $Z_P(\bh_w,\bh_u;l) = \mathop{\rm Tr}_{\bu,\bw} 
\prod_{\mu=1}^p \widehat{P}_{y_\mu}(u_\mu)
\prod_{i=1}^N P(w_i) \exp \left [
\bh_w \cdot \bw+\bh_u \cdot ({\rm i}\bu)+
({\rm i} \bu)^{\rm T}(lX) \bw \right ]$. 
This implies that the correct free energy (\ref{gibbs_FE}) can be 
obtained as $\Phi(\bm_w,\bm_u)=\widetilde{\Phi}(\bm_w,\bm_u;l=1)$
by setting $l=1$ in the generalized expression (\ref{gen_gibbs_FE}). 
One scheme to make use of this relation is to perform 
the Taylor expansion around $l=0$, for which $\widetilde{\Phi}(\bm_w,\bm_u;l)$
can be analytically calculated as an exceptional case, and substitute 
$l=1$ in the expression obtained, which is sometimes 
referred to as the Plefka expansion \cite{Plefka1982}. 
However, evaluation of higher order terms, 
which are not negligible for correlated patterns in general, 
requires a complicated calculation in this expansion, 
which sometimes prevents the scheme from being practically tractable. 
In order to avoid this difficulty, 
we take an alternative approach here, 
which is inspired by a derivative of equation (\ref{gen_gibbs_FE}) 
\begin{eqnarray}
\frac{\partial \widetilde{\Phi}(\bm_w,\bm_u;l) }{
\partial l}=-\left \langle ({\rm i} \bu)^T X \bw \right \rangle_l, 
\label{internal_energy}
\end{eqnarray}
where $\left \langle \cdots \right \rangle_l$ represents the 
average with respect to the generalized weight 
$\prod_{\mu=1}^p \widehat{P}_{y_\mu}(u_\mu)\times$
$\prod_{i=1}^N P(w_i)\times$ $\exp \left [
\bh_w \cdot \bw+\bh_u \cdot ({\rm i}\bu)+
({\rm i} \bu)^{\rm T}(lX) \bw \right ]$, 
$\bh_w$ and $\bh_u$ of which are determined so as to 
satisfy $\left \langle \bw \right \rangle_l=\bm_w$ and
$\left \langle ({\rm i}\bu) \right \rangle_l=\bm_u$, 
respectively \cite{OpperWinther2005}. 
The right hand side of this equation is an 
average of a quadratic form containing many random variables. 
The central limit theorem 
implies that such an average does not depend on details 
of the objective distribution but is determined 
only by the values of the first and second moments. 
In order to construct a simple approximation scheme, 
let us assume that the second moments are
characterized macroscopically by 
$\left \langle |\bw|^2 \right \rangle_l-
|\left \langle \bw \right \rangle_l|^2 =N \cw$
and $\left \langle |\bu|^2 \right \rangle_l-
|\left \langle \bu \right \rangle_l|^2 =p \cu$. 
Evaluating the right hand side of equation (\ref{internal_energy}) using 
a Gaussian distribution for which the first and 
second moments are constrained as 
$\left \langle \bw \right \rangle_l=\bm_w$, 
$\left \langle ({\rm i}\bu) \right \rangle_l=\bm_u$, 
$\left \langle |\bw|^2 \right \rangle_l-
|\left \langle \bw \right \rangle_l|^2 =N \cw$
and $\left \langle |\bu|^2 \right \rangle_l-
|\left \langle \bu \right \rangle_l|^2 =p \cu$, and 
integrating from $l=0$ to $l=1$ yields
\begin{eqnarray}
\widetilde{\Phi}(\cw,\cu,\bm_w,\bm_u;1)-\widetilde{\Phi}(\cw,\cu,\bm_w,\bm_u;0)
\simeq -\bm_u^{\rm T} X \bm_w -N F(\cw,\cu), 
\label{AdaTAP}
\end{eqnarray}
where the function $F(x,y)$ is provided as in equation (\ref{F_func})
by the eigenvalue 
spectrum of $X^{\rm T}X$, $\rho(\lambda)=N^{-1}
\sum_{k=1}^N \delta(\lambda-\lambda_k)$ and the 
macroscopic second moments $\cw$ and $\cu$ are included 
in arguments of the Gibbs free energy as the right hand side 
of equation (\ref{internal_energy}) depends on them. 
Utilizing this and evaluating $\widetilde{\Phi}(\cw,\cu,\bm_w,\bm_u;0)$, which 
is not computationally difficult since interaction terms are not included, 
yields an approximation of the Gibbs free energy as
\begin{eqnarray}
&&\Phi(\cw,\cu,\bm_w,\bm_u) \simeq -\bm_u^{\rm T} X \bm_w -N F(\cw,\cu) 
\cr
&&+\mathop{\rm Extr}_{\hcw,\bh_w}
\left \{\bh_w \cdot \bm_w-\frac{1}{2}\hcw \left (N \cw+|\bm_w|^2 \right )
-\sum_{i=1}^N \ln \left [\mathop{\rm Tr}_{w}
P(w) e^{-\frac{1}{2}\hcw w^2+h_{w i} w} \right ]
\right \} \cr
&&+\mathop{\rm Extr}_{\hcu,{\bh}_u}
\left \{\bh_u \cdot \bm_u
-\frac{1}{2}\hcu \left (p \cu-|\bm_u|^2 \right )
-\sum_{\mu=1}^p \ln \left [\int Dx
P(y_\mu|\sqrt{\hcu}x+h_{u \mu}) \right ]
\right \}, 
\label{TAP_free_energy}
\end{eqnarray}
which is a general expression of the {TAP free energy}
of the current objective system (\ref{posterior}). 
Extremization of this equation yields a set of {TAP equations}
\begin{eqnarray}
m_{wi}&=&\frac{\partial}{\partial h_{wi}}
\ln \left [\mathop{\rm Tr}_{w}
P(w) e^{-\frac{1}{2}\hcw w^2+h_{w i} w} \right ], \label{TAPw1}\\
\cw&=&\frac{1}{N}\sum_{i=1}^N \frac{\partial^2}{\partial h_{wi}^2}
\ln \left [\mathop{\rm Tr}_{w} 
P(w) e^{-\frac{1}{2}\hcw w^2+h_{w i} w} \right ], \label{TAPw2}\\
m_{u\mu} &=& \frac{\partial}{\partial h_{u\mu}}
\ln \left [\int Dx
P(y_\mu|\sqrt{\hcu}x+h_{u \mu}) \right ], \label{TAPu1}\\
\cu &=& -\frac{1}{p}\sum_{\mu=1}^p \frac{\partial^2}{\partial h_{u\mu}^2}
\ln \left [\int Dx
P(y_\mu|\sqrt{\hcu}x+h_{u \mu}) \right ],\label{TAPu2}
\end{eqnarray}
where
\begin{eqnarray}
\bh_w&=&X^{\rm T} \bm_u -2\frac{\partial }{\partial \cw} F(\cw,\cu)
\bm_w, \label{TAPhw1} \\
\hcw&=&-2\frac{\partial }{\partial \cw} F(\cw,\cu), \label{TAPhw2} \\
\bh_u&=&X \bm_w 
+\frac{2}{\alpha}\frac{\partial }{\partial \cu} F(\cw,\cu)\bm_u, 
\label{TAPhu1} \\
\hcu&=&-\frac{2}{\alpha}\frac{\partial }{\partial \cu} F(\cw,\cu),  
\label{TAPhu2}
\end{eqnarray}
solutions of which represent approximate values of the first and second moments
of the distribution (\ref{posterior}) for a fixed sample of $X$ (or $\xi^p$). 
$-2(\partial/\partial \cw) F(\cw,\cu) \bm_w$ and $ 
(2/\alpha) (\partial /\partial \cu )F(\cw,\cu)\bm_u$ 
in equations (\ref{TAPhw1}) and (\ref{TAPhu1}) 
are generally referred to as the {Onsager reaction terms}. 
Counterparts of these equations for systems of two-body 
interactions have been presented in an earlier article 
\cite{ParisiPotters1995}. 
Although we have assumed single macroscopic 
constraints as characterizing the second moments, the current 
formalism can be generalized to include component-wise multiple constraints
for constructing more accurate approximations, 
which leads to the adaptive TAP approach 
or, more generally, the expectation consistent approximate schemes 
developed by Opper and Winther \cite{OpperWinther2001L,OpperWinther2001,
OpperWinther2005}.

\section{Examples}
\subsection{Patterns of independently and identically distributed entries}
In order to investigate the relationship with existing results, 
let us first employ the developed methodologies to the case
in which the entries of $X$ are independently drawn from 
an identical distribution with zero mean and variance $N^{-1}$. 
This case is characterized by an eigenvalue spectrum 
of Mar\u{c}enko-Pastur type, 
$\rho(\lambda)=[1-\alpha]^+\delta(\lambda)+
(2 \pi)^{-1}\lambda^{-1}
\sqrt{[\lambda-\lambda_{-}]^+ [\lambda_{+}-\lambda]^+}$, 
where $[x]^+=x$ for $x>0$ and $0$, otherwise, 
and $\lambda_{\pm}=\left (\sqrt{\alpha}\pm 1 \right )^2$ 
\cite{TulinoVerdu2004}, which yields 
\begin{eqnarray}
F(x,y)=-\frac{\alpha }{2} xy. 
\label{IID_F_func}
\end{eqnarray}
This together with the relation 
$\left \langle \lambda \right \rangle_\rho=\alpha$, 
which holds for the current eigenvalue spectrum, 
implies that equation (\ref{A0}) can be expressed as
\begin{eqnarray}
\cA_0(\cw,\cu,\qw,\qu,m_w,m_u)
=-\frac{\alpha}{2} (\cw\cu + \qw\cu-\qu\cw-2 m_w m_u).
\label{IID_A0}
\end{eqnarray}
Inserting this into equation (\ref{replica_free_energy}) 
and then performing an 
extremization with respect to $\cu$, $\qu$ and $m_u$ yields
\begin{eqnarray}
\hcu=\cw, \quad \hqu= \qw, \quad \widehat{m}_u=m_w, 
\end{eqnarray}
where $\hcu$, $\hqu$ and $\widehat{m}_u$ are the variational variables 
used in equation (\ref{cAu}). 
This implies that the replica symmetric free energy (\ref{replica_free_energy})
can be expressed as
\begin{eqnarray}
&&\frac{1}{N}\left [\ln Z_P(\xi^p) \right ]_{\xi^p}
=\mathop{\rm Extr}_{\cw,\qw,m_w}\left \{
\cA_w(\cw,\qw,m_w) \right . \cr
&&\left .+ \alpha \mathop{\rm Tr}_y 
\int Dz Dx Q\left (y |
\sqrt{T_w-\frac{m_w^2}{q_w}}x+\frac{m_w}{\sqrt{q_w}} z \right )
\ln \left [
\int Dx P\left (y|\sqrt{\cw} x+\sqrt{\qw} z \right ) \right ] \right \}, 
\label{IID_replica_free_energy}
\end{eqnarray}
where the relation $\widehat{T}_u=\alpha^{-1}\left \langle 
\lambda \right \rangle_\rho T_w$ was utilized. 
This is equivalent to the general expression of the replica symmetric 
free energy of a single layer perceptron for pattern matrices with 
independently and identically distributed 
entries \cite{Engel2001,OpperKinzel1993}. 

\subsection{Gaussian linear vector channel}
The second example to show equivalent results 
to those obtained by earlier analysis 
is that of a Gaussian linear vector channel, which is 
characterized by $P(y|\Delta)=(2\pi \sigma^2)^{-1/2}$
$\exp \left [-(y-\Delta)^2/(2 \sigma^2) \right ]$ and 
$Q(y|\Delta)=(2\pi \sigma_0^2)^{-1/2}
\exp \left [-(y-\Delta)^2/(2 \sigma_0^2) \right ]$. 
In this case, equation (\ref{cAu}) is evaluated as
\begin{eqnarray}
\cA_u(\cu,\qu,m_u)=\frac{1}{2}\left (\sigma^2-\frac{1}{\cu} \right )\qu
-\frac{1}{2}\left (\sigma^2\cu -\ln \cu -1 \right )
-\frac{1}{2}\cu (\widehat{T}_u+\sigma_0^2 ),
\end{eqnarray}
while requiring that $m_u/\cu=1$. 
Further, extremization with respect to $\qu$ in equation 
(\ref{replica_free_energy}) indicates that $\Lambda_{\cu}$, which is 
the counterpart of $\Lambda_y$ in equation (\ref{F_func}) for $y=\cu$, 
is set to a constant value $\Lambda_{\cu}=\sigma^2$, 
which implies that $(\partial/\partial  \cu) F(\cw,\cu)
=(1/2) (\sigma^2-\cu^{-1})$
and $\cu=\sigma^{-2}+2 (\alpha \sigma^2)^{-1} \cw 
(\partial/\partial \cw) F(\cw,\cu)  $ hold.  
These, in conjunction with $\alpha \widehat{T}_u =T_w \left \langle \lambda 
\right \rangle_\rho$, 
indicate that equation (\ref{replica_free_energy}) can be
expressed as
\begin{eqnarray}
&&\frac{1}{N}\left [\ln Z_P(\xi^p) \right ]_{\xi^p}
=\mathop{\rm Extr}_{\cw,\qw,m_w} \left \{\cA_w(\cw,\qw,m_w) 
\right .\cr
&&\left . + G\left (-\frac{\cw}{\sigma^2} \right )+
\left (-\frac{T_w-2 m_w+\qw}{\sigma^2}+\frac{\sigma_0^2\cw}{\sigma^4} 
\right )G^\prime \left (-\frac{\cw}{\sigma^2} \right ) \right \}
-\frac{\alpha}{2}\left (\ln (2 \pi \sigma^2) + 
\frac{ \sigma^2_0}{\sigma^2} \right ), 
\label{G_vector_channels}
\end{eqnarray}
where 
\begin{eqnarray}
G(x)=\mathop{\rm Extr}_{\Lambda}\left \{
-\frac{1}{2}\left \langle \ln (\Lambda-\lambda) \right \rangle_\rho
+\frac{\Lambda}{2}x\right \}-\frac{1}{2}\ln x-\frac{1}{2}, 
\label{G_func}
\end{eqnarray}
is referred to as the Itzykson-Zuber integral or $G$-function 
in physics literature \cite{ItzyksonZuber1980,MarinariParisiRitort1994,
ParisiPotters1995, CherrierDeanLefevre2003}, 
which is linked to the $R$-transform 
of the cross-correlation matrix $X^{\rm T}X$
used in free probability theory \cite{TulinoVerdu2004,
VoiculescuDykemaNica1992,MullerGuoMoustakas2007}. 
Equation (\ref{G_vector_channels}) is equivalent to the expression 
for the replica symmetric free energy for Gaussian 
linear vector channels of a 
correlated channel matrix 
recently provided by the author and his colleagues 
\cite{TakedaUdaKabashima2006,TakedaHatabuKabashima2007}. 

\subsection{Ability of the Ising perceptron to separate random 
orthogonal patterns}
\begin{figure}[t]
\setlength{\unitlength}{1mm}
\begin{picture}(1870,80)
\put(15,-25){\includegraphics[width=110mm]{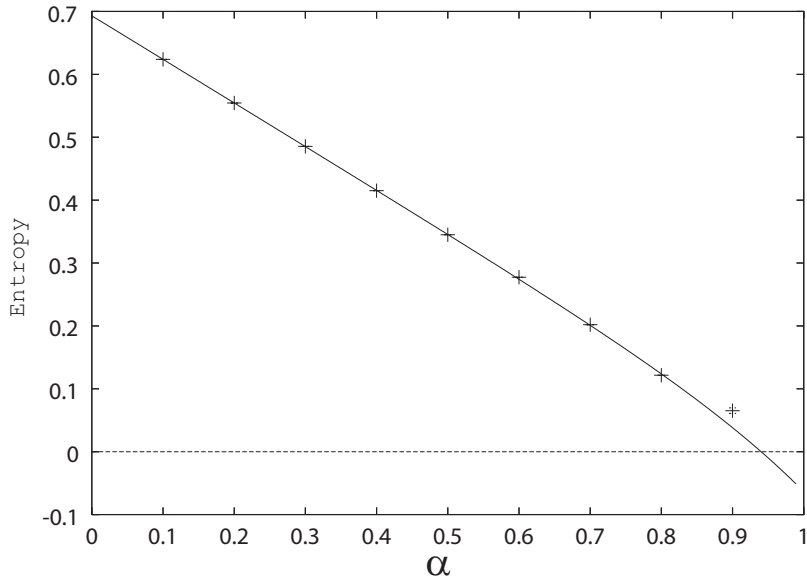}}
\end{picture}
\caption{Entropy of $\bw$ (per element) versus the pattern ratio $\alpha$.
For details, see the main text.}
\label{fig}
\end{figure}

In order to demonstrate the utility of the 
methodologies we have developed, as our final example we take up a simple but nontrivial 
problem concerning the separation ability of the Ising perceptron. 
Let us consider a simple perceptron of binary weight
$\bw =\{+1,-1\}^N$, $P(y|\Delta)=1$ for $y \Delta >0$ 
and 0, otherwise, where $y=\pm 1$. It is known that, 
in typical cases, this network can correctly separate 
a set of random patterns $\xi^p=\{(\bx_1,y_1),(\bx_2,y_2),
\ldots,(\bx_p,y_p)\}=(X,\by)$ up to $\alpha_c \simeq 0.833$, 
when the elements of $\bx_\mu $ are independently 
generated from an isotropic distribution 
and the elements of $y_\mu=\pm 1$ are independently and randomly
assigned with a probability of one half 
for $\mu=1,2,\ldots,p$ 
\cite{KrauthMezard1989,KrauthOpper1989,Derrida1991}. 
Our question here is how $\alpha_c$ is modified when 
the pattern matrix $X$ is generated randomly in such a way 
that the patterns $\bx_\mu$ are orthogonal to each other. 
In order to answer this question, we employ the replica and 
TAP methods developed in preceding sections 
for $\rho(\lambda)=(1-\alpha) \delta (\lambda)
+\alpha \delta(\lambda-1)$, which represents the eigenvalue spectrum 
of the random orthogonal patterns, assuming $0< \alpha <1$. 
Figure \ref{fig} shows how the entropy of $\bw$ depends on the 
pattern ratio $\alpha$. 
The curve indicates the theoretical prediction of the replica analysis
while the markers denote the averages of entropy 
obtained by the TAP method over 100 samples 
for $N=500$ systems. The error bars are smaller than the markers. 
Solutions of the TAP method are obtained 
by a method of iterative substitution, 
details of which are reported elsewhere \cite{ShinzatoKabashima2007}. 
Although the curve and the markers exhibit excellent agreement
for the data points $\alpha =0.1,0.2,\ldots, 0.8$, 
we were not able to obtain a reliable result for $\alpha = 0.9$, 
at which the iterative scheme does not 
converge in most cases even after 1000 iterations.  
This may be a consequence of RSB since the replica analysis indicates that 
the AT stability is broken at $\alpha_{\rm AT} \simeq 0.810$. 
Therefore $\alpha_c \simeq 0.940$ indicated by the condition 
of vanishing entropy is to be regarded not as the exact but as an approximate 
value provided by the unstable RS solution. However, extrapolation 
from the results of direct numerical experiments 
for finite size systems indicates that 
$\alpha_c \simeq 0.938$ \cite{ShinzatoKabashima2007}, 
which implies that the effect of RSB is not significant 
for the evaluation of $\alpha_c$ in this particular case. 

\section{Summary}
We have developed a framework for analyzing the inference performance
of densely-connected single-layer networks, typical examples of 
which are perceptrons and models of linear vector channels. 
The development is intended for dealing with correlated patterns. 
For this purpose, we have developed two methodologies based on 
the replica method and the Thouless-Anderson-Palmer approach, which are 
standard tools from the statistical mechanics of disordered systems, 
introducing a certain random assumption about the singular value 
decomposition of the pattern matrix. The validity and utility of the developed 
schemes are shown for two existing results and 
a novel problem. 

Investigation of the properties of algorithms for solving 
the TAP equations (\ref{TAPw1})--(\ref{TAPhu2}) 
\cite{Kabashima2003,MontanariPradhakarTse2005,MontanariTse2006}
and variants of them \cite{NeirottiSaad2005,TanakaOkada2005,BraunsteinZecchina2006}, as well as 
application of the developed framework to real world 
data analysis \cite{UdaKabashima2005,Weigt2007}
and various channel models \cite{Verdu1998,TulinoVerdu2004},
are promising topics for future research. 

\ack
This work was partially supported by Grants-in-Aid MEXT/JSPS, Japan, 
Nos. 1879006 and 17340116.  


\end{document}

%% file: commands.tex
\newcommand{\bTheta}{\mbox{\boldmath{$\Theta$}}}

\newcommand{\bm}{\mbox{\boldmath{$m$}}}
\newcommand{\bh}{\mbox{\boldmath{$h$}}}

\newcommand{\bw}{\mbox{\boldmath{$w$}}}
\newcommand{\bx}{\mbox{\boldmath{$x$}}}

\newcommand{\be}{\mbox{\boldmath{$e$}}}

\newcommand{\by}{\mbox{\boldmath{$y$}}}

\newcommand{\bu}{\mbox{\boldmath{$u$}}}

\newcommand{\mR}{\mathbb{R}}
\newcommand{\mN}{\mathbb{N}}
\newcommand{\qw}{q_w}
\newcommand{\qu}{q_u}
\newcommand{\hqw}{\hat{q}_w}
\newcommand{\hqu}{\hat{q}_u}
\newcommand{\cw}{\chi_w}
\newcommand{\cu}{\chi_u}
\newcommand{\hcw}{\hat{\chi}_w}
\newcommand{\hcu}{\hat{\chi}_u}
\newcommand{\bDelta}{\mbox{\boldmath{$\Delta$}}}
\newcommand{\cA}{{\cal A}}

\newcommand{\cqw}{{\cal Q}_w}
\newcommand{\cqu}{{\cal Q}_u}

%% file: inference_JPCS_final_rev2.bbl
\section*{References}
\begin{thebibliography}{99}
\bibitem{WatkinRauBiehl1992}
  {Watkin T L H, Rau A and Biehl M}
  {1993} {\it Rev. Mod. Phys.} {\bf 65} {499}

\bibitem{Engel2001}
  {Engel A and van den Broeck C} {2001}
  {\it Statistical Mechanics of Learning} (Cambridge: Cambridge University Press)

\bibitem{Nishimori2001}
  {Nishimori H} {2001}
  {\it Statistical Physics of Spin Glasses and Information Processing - An Introduction}
  (Oxford: Oxford University Press)


\bibitem{AdvancedMeanFieldMethods}
  {Opper M and Saad D (Eds.)} {2001}
  {\it Advanced Mean Field Methods: Theory and Practice}
  (Cambridge, MA: MIT Press)

\bibitem{MezardZeccina2002}
  {M\'{e}zard M, Parisi G and Zecchina R} {2002}
  {\it Science} {\bf 297} {812}

\bibitem{Tanaka2002}
  {Tanaka T}
  {2002} {\it IEEE Trans. on Infor. Theory} {\bf 48} {2888}

\bibitem{GuoVerdu2005}
  {Guo D and Verd\'{u} S}
  {2005} {\it IEEE Trans. on Infor. Theory} {\bf 51} {1983}

\bibitem{Muller2003}
  {M\"{u}ller R R}
  {2003} {\it IEEE Trans. on Signal Processing} {\bf 51} {2821}


\bibitem{Wenetal2005}
  {Wen C K, Lee Y N, Chen J T and Ting P}
  {2005} {\it IEEE Trans. on Signal Processing} {\bf 53} {2059}

\bibitem{Wen2006}
  {Wen C K, Ting P and Chen J T}
  {2006} {\it IEEE Trans. on Comm.} {\bf 54} {349}

\bibitem{Guo2006}
  {Guo D}
  {2006} {\it IEEE Trans. on Infor. Theory} {\bf 52} {1765}

\bibitem{Moustakas2003}
  {Moustakas A L}
  {2003} {\it IEEE Trans. on Infor. Theory} {\bf 49} {2545}

\bibitem{TakeuchiTanakaYano2007}
  {Takeuchi K, Tanaka T and Yano T} {2007} 
  {Asymptotic Analysis of General Multiuser Detectors 
    in MIMO DS-CDMA Channels } {\em Preprint} {arXiv:0706.3170}

\bibitem{Kabashima2003}  
  {Kabashima Y}
  {2003} \JPA {\bf 36} {11111}

\bibitem{NeirottiSaad2005}
  {Neirotti J P and Saad D}
  {2005} {\em Europhys. Lett.} {\bf 71} {866}

\bibitem{TanakaOkada2005}
  {Tanaka T and Okada M}
  {2005}
  {\it IEEE Trans. on Infor. Theory} {\bf 51} {700}

\bibitem{MontanariPradhakarTse2005}
  {Montanari A, Prabhakar B and Tse D} {2005} 
  {Belief Propagation Based Multi-User Detection } {\em Preprint}
  {arXiv:cs/0511044}

\bibitem{MontanariTse2006}
  {Montanari A and Tse D}
  {2006} {Analysis of Belief Propagation for Non-Linear Problems: 
    The Example of CDMA (or: How to Prove Tanaka's Formula)}
  {\em Proc. IEEE Inform. Theory Workshop} (Punta del Este: Uruguay)
  ({\em Preprint} arXiv:cs/0602028)

\bibitem{Verdu1998}
  {Verd\'{u} S} {\it Multiuser Detection}
  {1998}
  (Cambridge: Cambridge University Press)

\bibitem{TulinoVerdu2004}
  {Tulino A M and Verd\'{u} S} 
  {2004} {\it Random Matrix Theory and Wireless Communications}
  (Hanover, MA: Now Publishers)

\bibitem{TakedaUdaKabashima2006}
  {Takeda K, Uda S and Kabashima Y}
  {2006}
  {\it Europhys. Lett.} {\bf 76} {1193}

\bibitem{TakedaHatabuKabashima2007}
  {Takeda K, Hatabu A and Kabashima Y} {2007}
  {\it J. Phys. A: Math. Theor.} {\bf 40} {14085}

\bibitem{OpperWinther2001L}
  {Opper M and Winther O}
  {2001} \PRL {\bf 86} {3695}

\bibitem{OpperWinther2001}
  {Opper M and Winther O}
  {2001} \PR {\em E} {\bf 64} {056131}

\bibitem{OpperWinther2005}
  {Opper M and Winther O}
  {2005} {\em Journal of Machine Learning Research} {\bf 6} {2177}

\bibitem{Dotzenko2001}
  {Dotzenko V S}
  {2001} {\em Introduction to the Replica Theory of Disordered Statistical Systems}
  (Cambridge: Cambridge University Press) 

\bibitem{TAP1978}
  {Thouless D J, Anderson P W and Palmer R G}
  {1977} {\em Phil. Mag.} {\bf 35} {593}


\bibitem{ItzyksonZuber1980}
  {Itzykson C and Zuber J B}
  {1980} {\em J. Math. Phys.} {\bf 21} {411}

\bibitem{MarinariParisiRitort1994}
  {Marinari E, Parisi G and Ritort F} 
  {1994} \JPA {\bf 27} {7647}

\bibitem{ParisiPotters1995}
  {Parisi G and Potters M}
  {1995} \JPA {\bf 28} {5267}

\bibitem{CherrierDeanLefevre2003}
  {Cherrier R, Dean D S and Lef\`{e}vre A}
  {2003} \PR {\em E} {\bf 67} {046112}

\bibitem{GyorgyiTishby1990}
  {Gy\"{o}rgyi G and Tishby N}
  {1990} 
  {\em Neural Networks and Spin Glasses} ed Theumann W K and K\"{o}berle R 
  (Singapore: World Scientific) {p 3}

\bibitem{Tanaka2004}
  {Tanaka T}
  {2005} {\em Prog. Theor. Phys. Suppl.} {\bf 157} {176}

\bibitem{MPV1987}
  {M\'{e}zard M, Parisi G and Virasoro M A}
  {1987} {\em Spin Glass Theory and Beyond}
  (Singapore: World Scientific)

\bibitem{AT1978}
  {de Almeida J R L and Thouless D J}
  {1978} \JPA {\bf 11} {983}
 

\bibitem{Plefka1982}
  {Plefka T}
  {1982} \JPA {\bf 15} {1971}


\bibitem{OpperKinzel1993}
  {Opper M and Kinzel W}
  {1996} 
  {\em Models of Neural Networks III} ed Domany E, van Hemmen J L and 
  Schulten K (New York: Springer-Verlag New York) p 151



\bibitem{VoiculescuDykemaNica1992}
  {Voiculescu D V, Dykema K J and Nica A}
  {1992}
  {\it Free Random Variables} 
  (Providence, R.I.: American Mathematical Society)  

\bibitem{MullerGuoMoustakas2007}
  {M\"{u}ller R R, Guo D and Moustakas A L}
  {Vector Precoding for Wireless MIMO Systems: A Replica Analysis }
  {2007} {\em Preprint} {arXiv:0706.1169} 

\bibitem{KrauthMezard1989}
  {Krauth W and M\'{e}zard M}
  {1989} {\em J. Physique} {\bf 50} {3056}

\bibitem{KrauthOpper1989}
  {Krauth W and Opper M}
  {1989} \JPA {\bf 22} {L519}

\bibitem{Derrida1991}
  {Derrida B, Griffith R B and Pr\"{u}gel-Benett A}
  {1991} \JPA {\bf 24} {4907}

\bibitem{ShinzatoKabashima2007}
  {Shinzato T and Kabashima Y}
  {in preparation}

\bibitem{BraunsteinZecchina2006}
  {Braunstein A and Zecchina R}
  {2006} \PRL {\bf 96} 030201


\bibitem{UdaKabashima2005}
  {Uda S and Kabashima Y}
  {2005} {\em J. Phys. Soc. Jpn.} {\bf 74} 2233 

\bibitem{Weigt2007}
  {Braunstein A, Pagnani A, Weigt M and Zecchina R}
  {2007} {Gene-network inference by message passing} 
  {\em Proc. IW-SMI2007} (Kyoto) 137


\end{thebibliography}
